2011.04.06

# Astrometry history: Roemer and Gaia[1]

Erik Høg    Niels Bohr Institute

Copenhagen, Denmark

ABSTRACT: *During the Hipparcos mission in September 1992, I presented a concept for using direct imaging on CCDs in scanning mode in a new and very powerful astrometric satellite, Roemer. The Roemer concept with larger aperture telescopes for higher accuracy was developed by ESA and a mission was approved in 2000, expected to be a million times better than Hipparcos. The present name Gaia for the mission reminds of an interferometric option also studied in the period 1993-97, and the evolution of optics and detection in this period is the main subject of the present report. The transition from an interferometric GAIA to a large Roemer was made on 15 January 1998. It will be shown that without the interferometric GAIA option, ESA would hardly have selected astrometry for a Cornerstone study in 1997, and consequently we would not have had the Roemer/Gaia mission.*

## 1. Introduction

Only one astrometric satellite has been launched, Hipparcos, and its observations from 1989-93 brought a tiger leap of the accuracy and number of stars with good distances, proper motions and positions. In 2013, ESA will launch another large astrometric satellite Gaia, which is expected to bring a new tiger leap for astrometry. I have been deeply involved in both projects for 32 years from the very beginning of Hipparcos in 1975 when I made a completely new design of the satellite. I have written and lectured (Høg 2008, 2011) about these projects from my own perspective, but in a historically reliable manner with frequent checks of my memory by means of my archive and by correspondence with colleagues. I include personal recollections and reminiscences hoping to bring events and decisions closer.

In the summer of 1990 I began a collaboration with Russian colleagues about a successor for Hipparcos and we soon included Lennart Lindegren. This collaboration led to the Roemer proposal in 1992, to

GAIA in 1993, and then to Gaia and gradually more and more people contributed to the development. *The chain of ideas and actions related to optics and detectors is my main subject*, not at all a complete history of Gaia.

In the spring of 2010, I realized that the role of the Roemer satellite proposal of September 1992 (Høg 1993) seemed to be forgotten. This proposal was important for two reasons, many of the new ideas in the proposal are contained in the final Gaia satellite and the proposal in fact started the work towards Gaia. By October 1993, only a year after presentation of the Roemer proposal with direct imaging on CCDs, the basis had been laid for the studies of Roemer and another option, GAIA, using Fizeau interferometers (Lindegren et al. 1993b).

Interferometric designs were studied in the years 1993-97, followed by a design without any interferometry, but based on the ideas in Roemer with direct imaging on CCDs from full-aperture telescopes. The mission thus became a Roemer mission with large telescopes. The name GAIA with the capital "I" for interferometry remained, however, until about 2003 when it was changed to Gaia. The CCD as a two-dimensional detector with high detection efficiency is better by many orders of magnitude than the photoelectric detector in Hipparcos which measured only one star at a time and this potential advantage of a CCD was trivial by 1990 when my design of a new astrometric mission began, the only question was how to do it with CCDs.

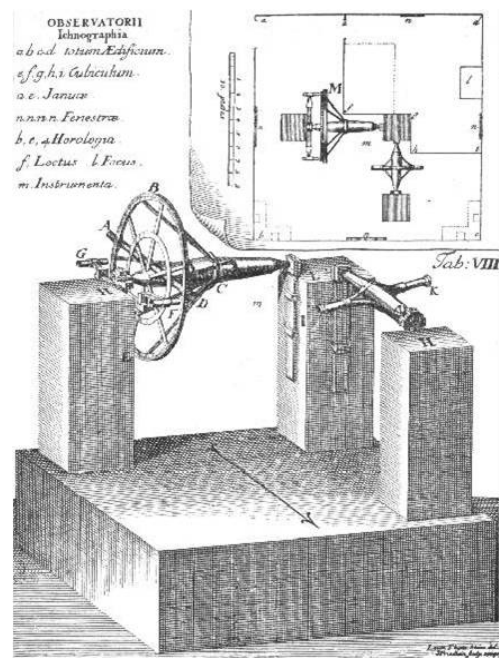

**Figure 1.** Ole Rømer (1644-1710) invented the meridian circle, the fundamental astrometric instrument for centuries.

---

[1] Quick link to other contributions to astrometry history are placed at the end of this paper.

I proposed the name "Roemer" as a proper name for an astrometric satellite. Ole Rømer (1644-1710) invented the meridian circle, see Fig. 1, the fundamental instrument of astrometry for centuries. He constructed many other physical and astronomical instruments in Denmark and he discovered the speed of light while observing in Paris.

Discussions and ensuing correspondence in 2010 led me to write about space astrometry plans in the 1990s, especially as related to Roemer and Gaia. I will begin with an account of what people are thinking or remembering of these things today, almost twenty years later. After an overview of Russian and American space astrometry follows the development of Roemer and Gaia in the 1990s. Then finally, very briefly, a view of the present Gaia design and of the future of space astrometry.

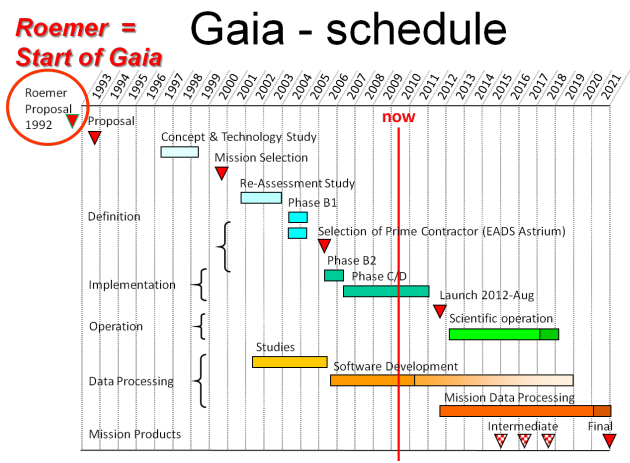

**Figure 2.** The Gaia schedule of mid 2010. I have inserted the time tag "Roemer proposal" at upper left.

## 2.  What People Thought in 2010

In the spring of 2010 I noticed that the time schedule for Gaia, Figure 2, began in late 1993. When I saw Timo Prusti, the Gaia Science Team leader, at the Gaia conference in Paris in June 2010 I suggested that he should recognise the role of Roemer by inserting a time tag at September 1992 with the words "Roemer proposal". This suggestion resulted immediately in remarks from several around the lunch table where we were seated that other suggestions had been made before 1992 saying that Hipparcos should be followed by another astrometric mission. I said that such suggestions are fine to make, but they did not trigger further work leading to Gaia. They were ideas or suggestions, not elaborate mission proposals as Roemer was.

Suggestions to use CCDs to measure hundreds of stars simultaneously had also been made it was continued at the lunch table. This is also an obvious idea, but not a mission proposal and this simple idea did not itself trigger the work that led to Gaia.

When I later mentioned this matter in a mail to several I received an offer from Francois Mignard. He proposed to include an image of Roemer on one of his slides if I wanted. This slide from his presentation in Paris as leader of the Gaia data reduction mentioned the "unfortunate followers" of Hipparcos aiming for 0.1 mas accuracy: "Roemer, FAME-1, FAME-2, DIVA, Lomonosov, AMEX". This was a kind offer by Francois, but not what I wanted. Here I want to stress that I do not blame Francois or anyone for not knowing so well what happened twenty years ago. How could I - when I see how much reading and correspondence was required to find out for myself. I want to thank Francois for many years of pleasant and efficient collaboration on astrometry and on the history of astronomy, including the present report.

Thus, in 2010 some participants in the Gaia preparations believed that merely general and vague ideas or suggestions had been made twenty years ago before the GAIA proposal and that the Roemer proposal was such a vague idea. This showed me that the history of Hipparcos-Roemer-Gaia and the relation to the many other mission proposals in the 1990s ought to be written. I contacted Michael Perryman, Lennart Lindegren and Ken Seidelmann and asked if they would collaborate in one way or another.

Michael answered immediately that he did not want to be involved, he felt no enthusiasm now about the history of Gaia. He continued: *"I did make my own extensive notes on the project, from its very beginnings, which cover the scientific process, the industrial design, the advisory committee politics, and many of the ESA internal issues. It runs (from memory) to some 40 pages of small text. … Perhaps, in years to come, I will write my own recollections of the first 10 years. But not now."* Lennart and Ken kindly sent extensive comments in the ensuing correspondence which have been used in the following.

## 3.  American Space Astrometry 1990

American astronomers in the U.S. Naval Observatory (USNO) and elsewhere have before 1990 been engaged in astrometry from space especially with the Hubble Space Telescope (e.g. Duncombe et al. 1990 and Seidelmann 1990) which has in fact provided accurate milliarcsecond astrometry, especially after the optical repair mission in 1993.

American astronomers also had plans for high-accuracy astrometry by Michelson interferometers in space with pointed telescopes, POINTS, see Chandler & Reasenberg 1990 and Figure 3.

In both cases this was narrow-field astrometry, very different from the global wide-angle astrometry provided by Hipparcos, Roemer and Gaia. The Americans thought of pointed telescopes in a satellite for observation of a few thousand stars with very high accuracy while ESA astronomers were developing a scanning satellite for systematic global astrometry of a hundred thousand stars with milliarcsecond accuracy.

The difference between the American approach to space astrometry and that within ESA was one reason that no cooperation has resulted which left significant traces in the early 1990s in either ESA or USA in spite of much communication at conferences and otherwise. Another reason was that the astrometric expertise in Europe was so sufficient that no collaboration on Hipparcos had been needed. But the inspiration to use CCDs in scanning mode came to me from America, from the work in those years on a meridian telescope in Arizona described by Stone & Monet 1990.

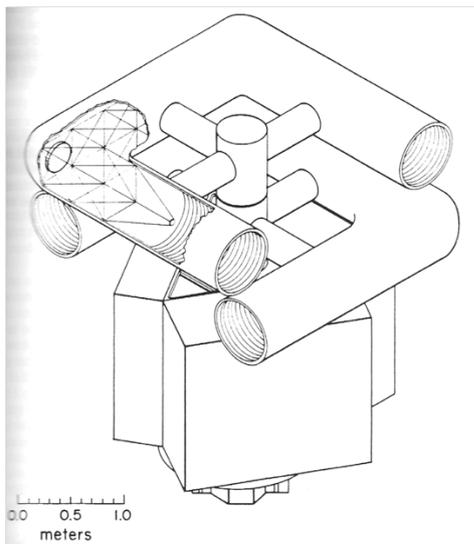

**Figure 3.** *"An artist's rendition of POINTS with 2-m separations between pairs of telescopes 25 cm diameter. The instrument, shown mounted on the Multimission Modular Spacecraft, comprises two U-shaped interferometers joined by a bearing that permits the angle between the principal axes of the interferometers to vary by up to a few degrees from its nominal value of 90 deg."* – quotation from original figure.

The idea from POINTS of Michelson interferometers in space for high-accuracy astrometry was further pursued in the 1990s and became the Space Interferometry Mission, also known as SIM Lite (formerly known as SIM PlanetQuest). It was a planned space telescope developed by the U.S. National Aeronautics and Space Administration (NASA), in conjunction with contractor Northrop Grumman. One of the main goals of the mission was the hunt for Earth-sized planets orbiting in the habitable zones of nearby stars other than the Sun. SIM was postponed several times and finally cancelled in 2010 (SIM 2011).

## 4. The Russian Collaboration

Inspiration to design a successor in space for Hipparcos could obviously not come to me from America. Nor could it come from the Hipparcos community where all attention was focused on the Hipparcos observations and data reduction. The inspiration came from Russia as I have described before (Høg 2007), to be partly repeated here with new details.

At a conference in Leningrad in 1989 we had heard about three plans in Russia, then USSR, for successors to Hipparcos. This became crucial for the development of Roemer and Gaia because I met an active interest in Leningrad and Moscow during the following years, especially with Mark Chubey and his team in the Pulkovo Observatory and with the Mission Control Centre in Moscow, without which there would have been no Roemer or Gaia mission today.

The three Russian plans were described at the IAU Symposium No. 141 held in Leningrad in October 1989: (1) Lomonossov with a pointing telescope of 1 m Ø, F=50 m aiming for 1 mas accuracy. (2) REGATTA-ASTRO: scanning telescope, 10 mas accuracy. (3) AIST shown in Figure 4: 2 telescopes, 0.25 m Ø, scanning, 1 mas.

All three aimed for launch before 1997 and Lomonossov and AIST expected 1 mas accuracy, similar to Hipparcos. The proposers considered as the primary scientific aim to get second epoch positions and thus very accurate proper motions for the 100 000 Hipparcos stars.

In fact, the inspiration to design a new mission came for me at a visit in the summer of 1990 to the Caucasus mountains with Mark Chubey and his team (see Figure 6), after I had lectured about Hipparcos and Tycho in Pulkovo and Moscow.

At first, during our discussions on the travel I just wanted to understand the Russian projects, especially the AIST, see Chubey, Makarov, Yershov et al. 1990 and Figure 4. But after a day's discussion I realized that I was thinking more about improving Hipparcos than about understanding how AIST was supposed to work.

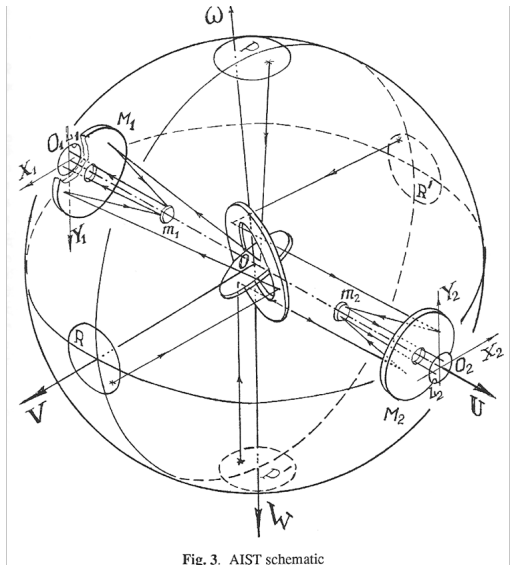

**Figure 4.** The proposed Russian astrometric satellite, AIST, with two telescopes.

Staying at the Kislovodsk Observatory in the Caucasus, I went out of bed in the first night and started to put my thoughts into a first drawing. I thought I was quiet, but Mark had heard me from his room and knocked my window asking in a concerned voice whether I was ill which I could truly deny. He then arranged for a better lamp at my table.

That was the morning of the first day with clear weather so that I could see the beautiful snow covered double peak of Mount Elbrus, an extinct volcano and the highest mountain in all of Europe. The day before they had sometimes pointed with the arm into the clouds saying: "Elbrus is over there". Now I understood why it was so important for them to point.

Our discussions were continued and one of the following meetings took place in Moscow in June 1991, now also with Lennart Lindegren present. It was important to involve Lennart in the design, he could make the correct estimate of the astrometric accuracy, always unfailing in mathematics.

I reported in the Hipparcos Science Team after every meeting with the Russians. Our leader Michael Perryman was a bit reluctant to give time, but he usually gave the 10 or 15 minutes I wanted, and useful discussion resulted. Michael was of course reluctant because it was his task as a leader to look after the observations and data reduction of the flying satellite Hipparcos, not to design a new mission, but he became very interested and active in such design in 1993.

At the Moscow meeting in June 1991 we presented ideas for a second Hipparcos, see Figure 5. The Hipparcos system, still with IDTs and photomultipliers, was improved with larger telescopes and enhanced star mappers. Expected accuracy was 1 mas for 400 000 stars and 0.2 mas/year proper motions for the 120 000 Hipparcos stars. We wrote: "The proposal … is being considered by the Mission Control Centre, Moscow."

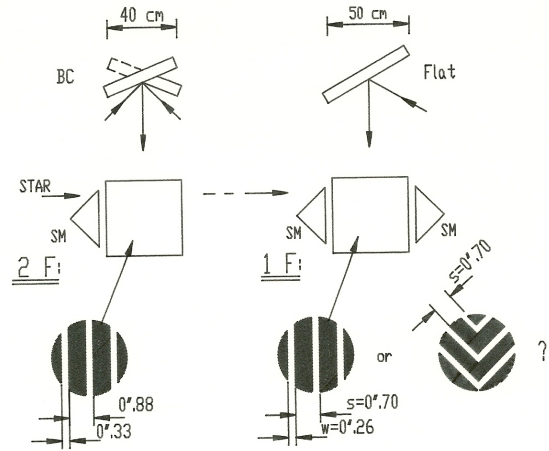

**Figure 5.** A second Hipparcos with two telescopes, proposed in 1991 by Høg & Chubey. The 2F telescope observes two fields by means of a beam combiner. The 1F telescope with larger aperture has only one field of view. The detectors are still photoelectric as in Hipparcos.

The paper was accepted for publication, but the proceedings never appeared. It was also presented as a poster at the IAU General Assembly in Buenos Aires in August 1991, but was not accepted for publication. It is now scanned and placed on my homepage, Høg & Chubey 1991.

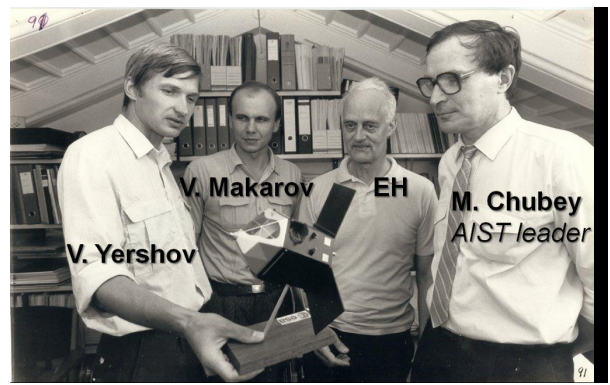

**Figure 6.** The Russian space astrometry team visited Copenhagen in 1991 and is here looking at the Hipparcos model. One member, Valeri Makarov, then stayed seven years in Copenhagen working on the Tycho-1 and Tycho-2 Catalogues.

After this exercise my study began of how to use a CCD (Charge Coupled Device). This two-dimensional detector was invented in 1969 by W. Boyle and G.E. Smith who subsequently received the Nobel Prize for physics in 2009. In 1979 an RCA 320x512 pixel cooled

CCD system was first used on a 1-meter telescope at Kitt Peak National Observatory. This started the take-over from photographic plates in astronomy, including astrometry, and a similar take-over from films in hand-held digital cameras began ten years later.

Two proposals were made for the use of 2D detectors in Hipparcos according to mails in 2011 from M. Perryman and R. Le Poole. A proposal to use a solid-state two-dimensional detector, an ICID – not a CCD, in Hipparcos was made by di Serego Alighieri et al. 1980. Another proposal by M. Hammerschlag in 1981 transported the charges in a CCD perpendicular to the motion of the stars, while Gaia transports along with the motion of the stars in TDI mode (Time Delayed Integration). By 1990 when my work towards Roemer and Gaia began the potential advantage of using a 2D solid-state detector was trivial, the only question was how to do it. The two proposals were in fact not known to me at that time.

The higher quantum efficiency of a CCD and the ability to observe many stars simultaneously would be the great and very obvious advantage over the photoelectric detectors. I learnt in 1991 from our electronics engineer in Copenhagen, Ralph Florentin Nielsen, what a CCD can do and what it cannot do.

But there was doubt in those years about the use of CCDs for astrometry. Their dimensional stability was doubted, the sensitivity was perhaps not uniform over the individual pixels, and the position of the pixels was perhaps not stable and had to be calibrated.

With these concerns in mind I began with a design using the CCD as a modulation detector. We knew from Hipparcos that a very accurate grid could be manufactured and would be very stable, so that seemed to be the way to go and out came the design in Figure 7. We could see that the astrometric efficiency of the satellite would be 1000 times higher than that of Hipparcos, as always based on Lennart's calculations.

## 5. Roemer and GAIA 1992-1994

The proposal with a CCD as modulation detector was called Hipparcos-2 in a report of January 1992. It was submitted to the IAU Symposium No. 156 which was to be held in September 1992 in Shanghai. But in May my study began of direct imaging on CCDs and this soon promised to be a hundred times better than the system with modulation so I thought it deserved the name Roemer, not just Hipparcos-2. We brought both manuscripts to Shanghai and both were accepted for oral presentation (Høg & Lindegren 1993, and Høg 1993, respectively.)

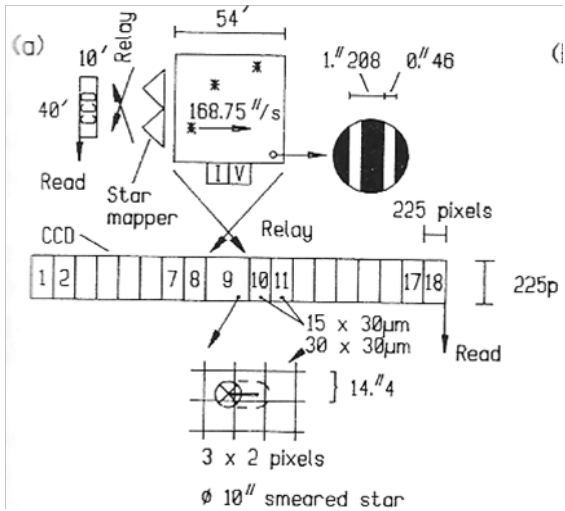

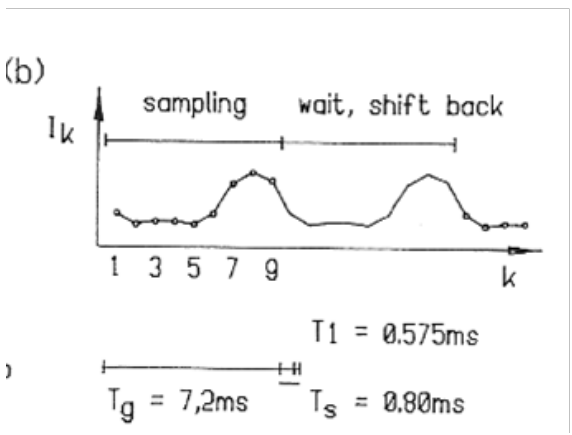

**Figure 7.** The detection of modulation with a CCD was proposed in 1992 for a Hipparcos-2 satellite (Høg & Lindegren 1993).

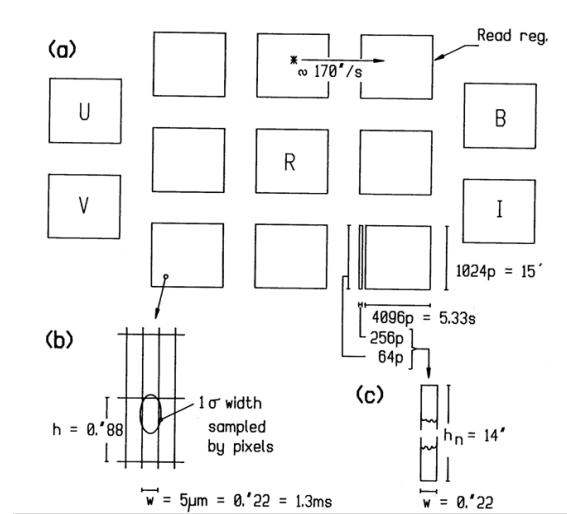

**Figure 8.** Focal plane of the Roemer satellite with CCDs in scanning mode proposed in 1992, stars moving left to right through the field (Høg 1993).

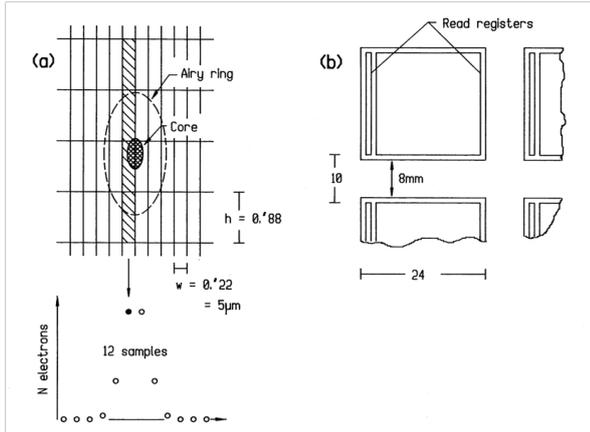

**Figure 9.** Sampling of the Roemer CCDs. (a): The pixels are elongated perpendicular to the scanning, several pixels (here 4) are read together in order to decrease the readnoise, and (b): short CCDs are provided for bright stars, features later adopted in Gaia.

Roemer was a very specific mission proposal with CCDs in time delayed integration and with direct imaging of the stars on the CCDs (Høg 1993, see Figures 8 and 9). It was a scanning satellite with a beam combiner similar to that in Hipparcos. For a 5 year mission an astrometric accuracy of 0.1 mas was predicted at V=12 mag, more than 10 times better than Hipparcos. The astrometric efficiency was 100 000 times higher, but obtained with the same telescopic aperture of 29 cm as Hipparcos. Astrometry and multicolour photometry for 400 million stars were included.

The use of a CCD directly in the focal plane is astrometrically much more efficient than a modulating grid as in Hipparcos because much light is lost in the grid. The use of many CCDs with their higher quantum efficiency than the photoelectric IDT detector of Hipparcos, smaller transmission losses than in the IDT relay system, and the capability to observe thousands of stars simultaneously translates into at least 100 000 times higher astrometric efficiency for the same telescope aperture.

Further improvement of accuracy through larger aperture would come from the higher angular resolution and from the larger number of photons collected. This path was chosen for the Gaia mission development after the studies from 1993-1997 had shown that interferometry was not the way to go, as shall be elaborated below.

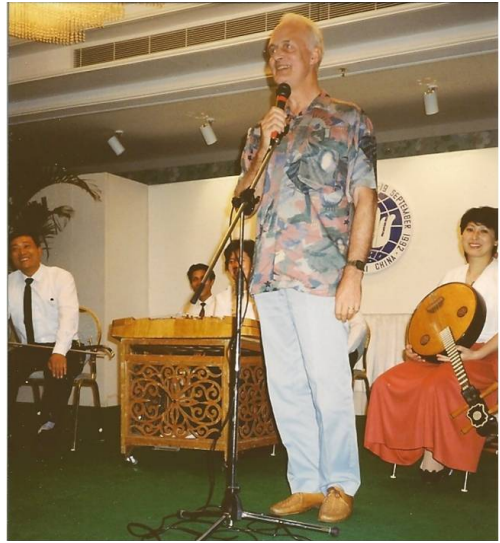

**Figure 10.** Shanghai 1992. You are encouraged to entertain at the conference dinner, I did so in Chinese and other languages.

During the conference in Shanghai three of us, Jean Kovalevsky, Ken Seidelmann and myself, agreed to apply for an IAU Symposium dedicated to sub-milliarcsecond optical astrometry. The symposium was approved and held in The Hague, August 15-19, 1994 (Høg & Seidelmann (eds.) 1995) and we wrote in the preface to the proceedings: *"Astrometry is on the threshold of great changes due to the fact that this decade, alone, is witnessing an improvement of stellar positions equivalent to the total improvement of the previous two centuries."*

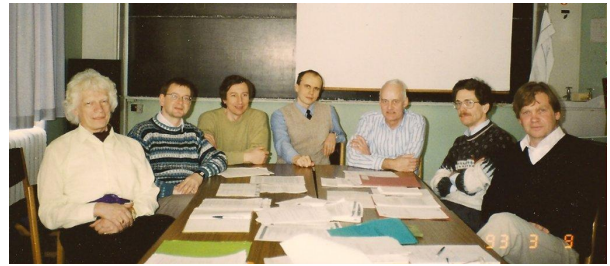

**Figure 11.** Roemer proposers met in March 1993 in Copenhagen: Kovalevsky, Lindegren, Halbwachs, Makarov, Høg, van Leeuwen, Knude – missing here: Bastian, Gilmore, Labeyrie, Pel, Schrijver, Stabell, Thejll.

The ideas in Roemer were adopted in a mission proposal submitted to ESA on 24$^{th}$ May 1993 for the Third Medium Size ESA Mission (M3). We proposed to measure 100 million stars and to obtain an accuracy of 0.2 mas at V=13 in a 2.5 year mission with a 34 cm telescopic aperture. Some of the proposers were members of the Hipparcos Science Team (see Figure 11 and Lindegren et al. 1993a).

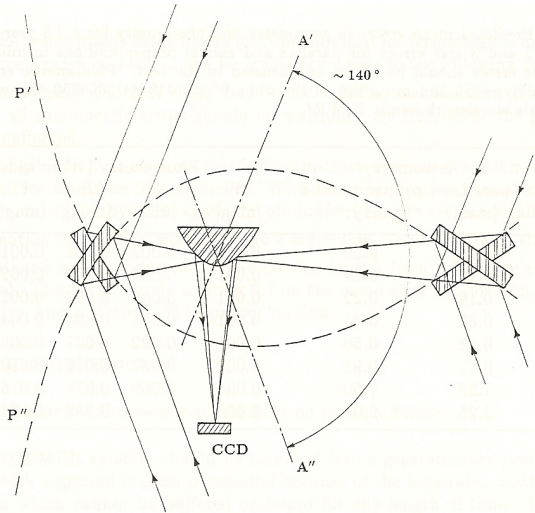

**Figure 12.** *"Optical sketch of a Fizeau-type scanning interferometer. The beam combiners are part of confocal paraboloids (P', P'') whose axes A' and A'' make a fixed angle of approx 140 deg."* – quotation from the original figure of 1993.

The mission was in fact called ROEMER, an acronym for "Rotating Optical Observatory for Extreme Measuring Efficiency and Rigour", but we will here simply call it Roemer. A section called "The FIZEAU option", not part of the baseline proposal, was included *"to point out a possible improvement towards a scanning satellite with ten times the angular accuracy of ROEMER"*. The section described a scanning satellite with *"two confocal Fizeau-type (or 'wide field') interferometers whose axes form a basic angle of the order 140 deg."*, a description fitting very well to the later GAIA. But the included optical system, shown here in Figure 12, underwent major development before it was called GAIA, an acronym for Global Astrometric Interferometer for Astrophysics.

The Roemer proposal was also presented at a conference in Cambridge in June 1993 (Høg and Lindegren 1994).

The proposal to ESA was rated by the Astronomy Working Group (AWG) to be the best among all astronomical proposals for M3. But it was considered to come too soon after Hipparcos and it was not sufficiently ambitious with respect to accuracy. It was therefore referred to a Cornerstone Mission study if 10-20 µarsecond accuracy could be demonstrated.

The proposal of a Cornerstone study meant that the AWG members got rid of a competitor for the M3 mission, but we should be grateful that we did not get approval for M3 since that would have prevented us ever to design the much more powerful Gaia.

As a reply to an ESA call for proposals of Cornerstone studies we submitted on 12 October 1993 a proposal to study for astrometry "a large Roemer option and an interferometric option", GAIA. They should be studied as two concepts for an ESA Cornerstone Mission for astrometry "without a priori excluding either", as Lindegren wrote in the cover letter.

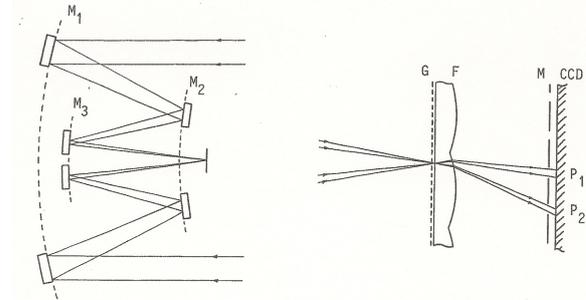

**Figure 13.** The GAIA system as it appeared in October 1993 (Lindegren et al. 1993b). Fizeau interferometer at left, modulating grid (G) with field lenslets (F) and detector (CCD).

The interferometric concept based on the above FIZEAU option was included in the report by Lindegren et al. 1993b, a concept which was mainly a result of discussions between Lindegren and Perryman. Two features in Figure 13, compared with Figure 12 should be mentioned. Firstly, the parabolic mirrors, which would give only a very small usable field, have been replaced with a three-mirror telescope, and secondly, principles of a CCD detection system are indicated.

In September 2010 Lindegren wrote to me: *"... in April 1994 during the HST meeting in Lund (14-15 April), I did write (in consultation with Michael) an e-mail to Steven Beckwith, the chairman of the UV-to-radio topical team of the Survey Committee then drafting ESA's Horizon 2000+ plan. In the e-mail, which was copied to L. Woltjer (chairman of the Survey Committee),* **I again stressed that Roemer and GAIA should not be seen as competing projects** *but as an indication of the different ideas circulating in the community, and the strong conviction that an advanced astrometric mission would be technically feasible and extremely worthwhile",* quoted from Lindegren 1994.

Development of the ideas was continued, especially by Lennart Lindegren, Michael Perryman and myself, of the proposed two mission concepts with higher accuracies: First, the interferometric mission GAIA, (see Figure 14) and second, a Roemer mission, called Roemer+, (see Figure 15) with larger apertures by Høg in August 1994 (Høg 1995).

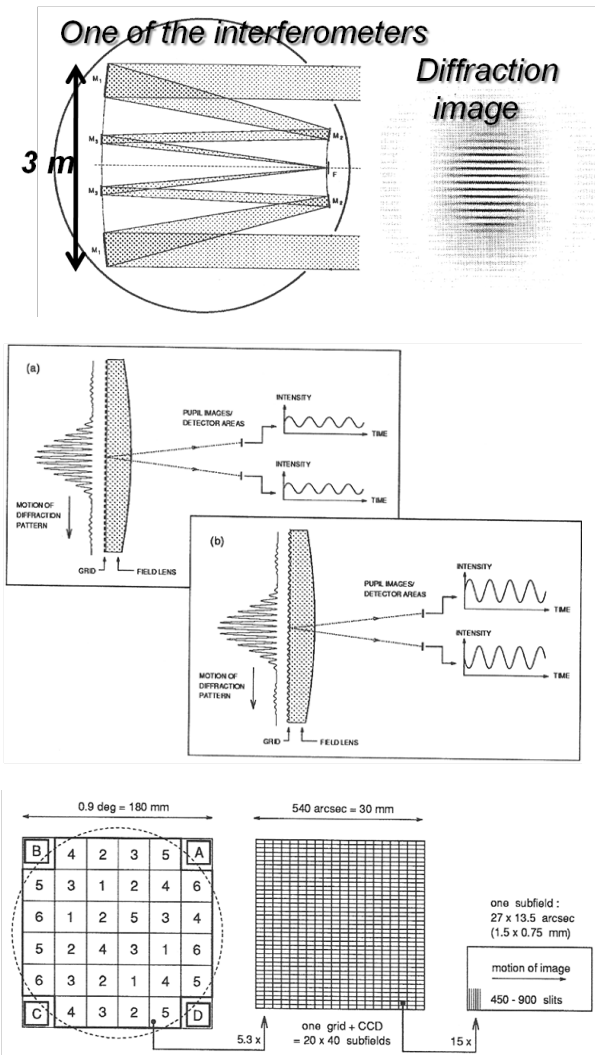

**Figure 14.** The GAIA system as it appeared in August 1994 (Lindegren & Perryman 1995). The optical system at top, then detection of the modulation, and more details below.

Thus, the basis for the studies of Roemer and GAIA was laid within one year and expanded with the large Roemer+ in August 1994 which could reach the 10-20 µarcsec goal. But by 2010 the role of Roemer had been forgotten, although the present Gaia may be seen as a large Roemer.

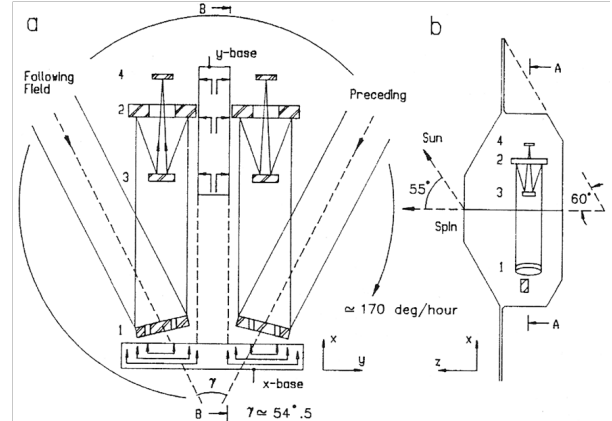

*Figure 2.* The Roemer+ satellite design. (a) Two Baker-Schmidt telescopes with tilted reflective corrector plates are pointed at the scanning great circle. The optical components are monitored by interferometric distance gauges with picometer precision, thus obtaining the variations of the basic angle as function of time. (b) Section of the rotationally symmetric satellite.

**Figure 15.** The Roemer+ satellite design of August 1994 (Høg 1995a), the first large Roemer. The use of picometer sensors is indicated.

## 6. Studies of interferometry 1993-1997

Ken Seidelmann writes to me on March 10, 1994 that USNO is pursuing NEWCOMB which is however unfunded, and that they are interested in collaboration on other space-based astrometry projects like Roemer and GAIA.

In August 1994 American astronomers presented the Newcomb Astrometric Satellite, *"a concept for a small, quick, inexpensive, initial optical interferometer in space"*. It *"would have a stacked set of 3, or 4, Michelson optical interferometers..."*. It would be a pointing satellite with a precision of 0.1 milliarcsecond. Requirements, but no specific design was included." Quotations are from Johnston et al. 1995.

In early 1995 they began to think about wide angle astrometry with a scanning satellite. They proposed FAME with Fizeau interferometers as a MIDEX mission of NASA (Johnston 1995a and 1995b), but did not succeed. This proposal is called FAME-1 or the "first FAME design" in the following to distinguish from the FAME-2 proposal a few years later.

In July 2010 Ken Seidelmann wrote: *"Ken Johnston and I started with the Newcomb proposal which evolved into the first FAME design, which was a Fizeau interferometer with a JPL optical design. That proposal was submitted to NASA. I gave a presentation in Europe on that design…"* It was in Cambridge in June 1995 (Seidelmann et al. 1995).

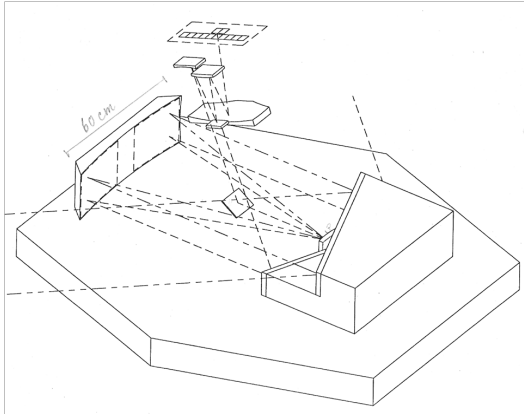

**Figure 16.** The FAME-1 optical design with Fizeau interferometer of September 1995.

I was a member of the FAME team from May to about November 1995 on invitation by Ken Johnston, but my duties did not allow me to continue. In October 1995 Johnston proposed to visit Copenhagen to meet with Lennart, Floor, and me to discuss the FAME proposal, but we answered in November that we were too pressed for time since Hipparcos and Tycho were still occupying us. Likewise we were too busy to offer a collaboration on the data reduction.

In September I received an improved optical system shown in Figure 16, also included in Johnston 1995a. That led me to design a new system, GAIA95, in October 1995, also a Fizeau interferometer but built into a Gregorian telescope, see Figure 17. A prism placed at the intermediate exit pupil lets the light through a hole in its middle. The focused but almost parallel beam returned from the secondary S3 passes through the prism thus providing spectra of all stars perpendicular to the scan direction. Astrometry and photometry could be obtained from the same images, the dispersed fringes. The system provided imaging without any disturbing central obscuration, utilizing the space between the two beams of the Fizeau interferometer.

The system was studied in Copenhagen and a report was distributed soon after by Høg, Fabricius & Makarov 1995. The publication by the same authors appeared in 1997.

The GAIA95 system with a D=1.5 m primary is shown in Figure 17. It could provide 10 µarcsec precision at V=14 mag and was considered for GAIA in 1997.

Soon after the first report had been distributed in 1995 I was called by Uli Bastian from Heidelberg. He asked if I could keep a secret for a few months. I promised and he continued saying that I could look forward to a small satellite on the GAIA95 design being launched before my 70[th] birthday - which would be 17 June 2002.

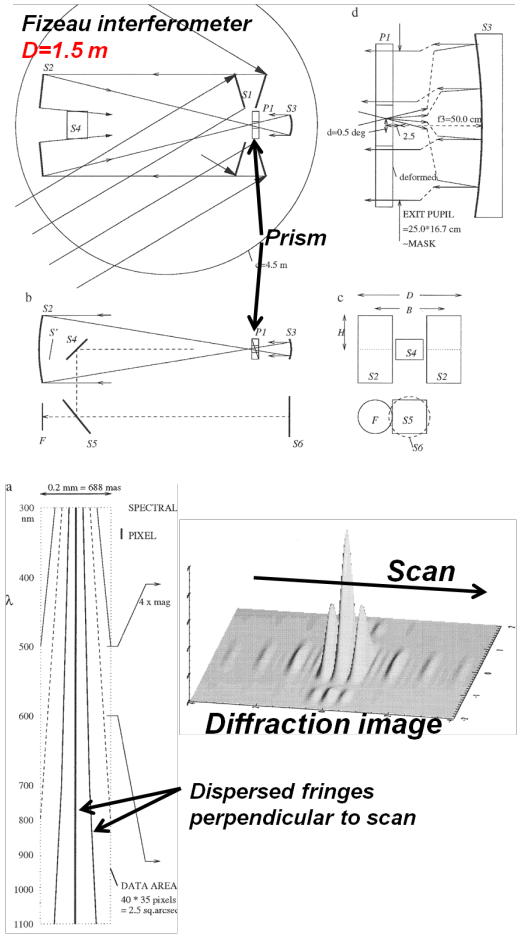

**Figure 17.** The GAIA95 design of October 1995 by Høg, Fabricius & Makarov, a Fizeau interferometer with Gregorian telescope, providing dispersed fringes for simultaneous astrometry and spectrophotometry.

Uli and German colleagues were working on a project which became the German interferometric minisatellite for astrometry and photometry, DIVA, Röser et al. 1996. The DIVA telescope was scaled to one tenth of GAIA95, i.e. to an aperture D=15 cm. In the fall of 2000 it was officially selected as a minisatellite project by the German space agency DLR. It was abandoned in 2002 when one of the German funding partners had dropped out. From 2002 to 2003 a follow-up collaboration with the USNO was tried, partly under the name of AMEX. This effort ended in 2004 when NASA funding did not materialise, and German astronomers decided to focus on the GAIA mission.

Ken Seidelmann wrote recently that many of the considerations concerning Gaia were going on at roughly the same time for the scanning satellite FAME2 (Johnston 2003), e.g. was the interferometer dropped in 1997 or 98. It was approved as a NASA MIDEX project in 1999, but was cancelled in 2002 primarily due to CCD chip production problems and budget concerns.

But these developments do not belong to our subject. The present report is about the development of the various optics and detector systems of Roemer, GAIA, and Gaia, including the scientific and technical environment for a better understanding. It is not a history of the Gaia project, nor is it of course a history of space astrometry in total. An overview of "future astrometric missions" is given in August 1996 by Seidelmann (1998). Seidelmann briefly describes ten plans (one by ESA, one in Germany, one in Japan, three in Russia, and four in the USA) of which one has survived, the interferometric GAIA, *"with a possible launch date of 2015."*

I was a team member of DIVA at the beginning, but then concentrated on the GAIA development. Michael Perryman insisted, and rightly so, that a member of the GAIA team could not also be member of another space astrometry team. The GAIA team had the task of developing a project for ESA, and should not be a centre for development of astrometric satellites.

The intense work on space astrometry development during these years appears from the listed proceedings of four international astrometric meetings: 1993 in Cambridge, UK, 1994 in The Hague, 1995 again in Cambridge, and 1997 in Venice. References are given below to the 22 papers especially on the Roemer and GAIA optics and detection systems. We find only one paper of this kind in 1993, two in 1994, 13 in 1995 and six in 1997. The GAIA95 option is only one example from the development of optics and detection for the interferometric GAIA in those years, but rather interesting because of the connection to American space astrometry and to DIVA.

Among the 22 papers, only four discuss the non-interferometric full-aperture Roemer option (Høg & Lindegren 1994, Høg 1995a, and 1995d, and Yershov 1995.) We scientists did not follow Lindegren's recommendation to ESA in 1993 and 1994 to study the Roemer and GAIA options "without a priori excluding either".

We were all very fascinated by the idea of Fizeau interferometry and we worked hard on its development though still occupied primarily with the data reduction and publication of Hipparcos and Tycho results. But our opinion changed when the ESA Cornerstone study started and the ESA Science Advisory Group, led by Michael Perryman, could discuss the studies by the industrial and ESA teams. We saw then that interferometry was the wrong track and we returned to direct imaging on CCDs in telescopes with apertures as large as could possibly be contained in the ESA launcher.

## 7. Cornerstone study approved 1997

The capital letter "I" in GAIA stood for Interferometry since the proposal in 1993 and the name was maintained although interferometry was dropped already in January 1998, but about 2003 the name was changed from GAIA to Gaia. In 2007 I proposed to change the name from Gaia to Roemer (Høg 2007) in recognition of Ole Rømer, this proposal of course came too late.

The names GAIA or Gaia have been maintained throughout the years for the sake of continuity and because we did not want to draw attention to the fact that interferometry had been dropped since we knew that ESA had attached importance to the use of interferometry when the Cornerstone Mission study was approved.

This appears from the letter of invitation to join the Science Advisory Group (SAG) for an Astrometry Cornerstone. The letter of invitation (ESA 1997) is dated 11 March 1997 and reads:

*"Space interferometry was identified in the ESA long-term programme for space science, Horizon 2000, as a potential candidate among space projects planned for after the turn of the century.*

*Recently, a Survey Committee established by ESA, has updated this programme as the Horizon 2000 Plus plan, which identifies three major projects over the period 2006-2016. One of these is an interferometry observatory as a Cornerstone mission open to the wide scientific Community. The first option would be to perform astrometric observations at the 10 µarcsec level. As an alternative option, the Survey Committee recommends studies of infrared interferometry, in particular with the aim of detecting planets around other stars. ... in January 1997, ESA's Space Science Advisory Committee recommended to start the study activities in preparation for the future definition of these interferometry projects. ..."*

The Cornerstone study of infrared interferometry mentioned led to the project Darwin. On the ESA Portal I now looked for "interferometer" and found a page from August 1997 about Cornerstones for GAIA, Darwin, and LISA. Asking for Darwin the answer was a page beginning with *"Study ended 2007, no further activities planned"*.

It appears that astrometry by interferometry is mentioned in the invitation, but neither GAIA nor Roemer. It seems therefore, that astrometry would not have been chosen by the Survey Committee for a Cornerstone study if our proposal had only contained a large Roemer and if no interferometry had been included. We were cautious in the Science Group never to make a point of the fact that interferometry was dropped after less than a year of study by industry. We continued with the Roemer option reasoning that ESA wanted the astrometric science and not a particular instrument.

But today we should lay open what really happened. (1) Without the Roemer proposal of 1992 there would have been no GAIA proposal in 1993, (2) without the GAIA with interferometry no selection for a Cornerstone study in 1997, and (3) without the Roemer concept of direct imaging with full-aperture telescopes there would not have been a feasible astrometry mission to approve in 2000 since the approved GAIA or Gaia may be seen as a large Roemer mission with many of the features proposed for Roemer in 1992.

It would be interesting to know what happened in the Survey Committee and the SSAC in order to understand why they selected interferometry. The minutes of the meeting must still exist and some participants could be interviewed. It seems clear that they were not asking for the best possible astrometry because they ignored the clear recommendation of the astrometry experts in the proposal by Lindegren et al. 1993b and repeated in the letter by Lindegren 1994. The recommendation was to study *"a large Roemer option and an interferometric option ... without a priori excluding either."* The ESA committees had a "great vision of interferometry" rather than a vision of great astrometry.

In an ESA committee of astronomers most members will be astrophysicists. They will often consider astrometry to be very useful for astronomy, but when it comes to a decision between expensive projects, astrometry has a very difficult standing. This was the case at the approval of Hipparcos as I have shown in Høg 2011a: *"Miraculous approval of Hipparcos in 1980"*. Several miracles happened then. Miracles only happen when good and strong persons take action.

### 8. From GAIA to Roemer/Gaia

The approval of a 'Concept and Technology Study' for GAIA (along with the other three cornerstone mission candidates) was given in 1996 and an ad hoc Science Advisory Group (SAG) was established in March 1997. *"A one-year industrial study took place between mid-1997 and mid-1998. Three industrial proposals were submitted in June 1997. The contract subsequently was awarded to Matra-Marconi Space (MMS) in July 1997"*, quotation from ESA 2000.

The GAIA SAG had its first meeting in March 1997 led by M. Perryman and the members were: F. Mignard, P.T. de Zeeuw, G. Gilmore, E. Høg, M. Lattanzi, L. Lindegren, and S. Röser – K.S. de Boer and X. Luri joined the SAG later. The following three years of work resulted in the Concept and Technology Study ESA 2000 which presents the scientific case of GAIA on 100 pages and the technical design, mission performance, and data analysis on a further 300 pages.

From these years of intense work I shall here only mention some of the main steps in the design of the payload. I will describe only few individual contributions, but I want to emphasize that Michael Perryman was our very efficient and competent leader all the time – without Michael Perryman and Lennart Lindegren there would be no Gaia.

The first design, Figure 18, from June 1997 corresponds to the GAIA proposal by Lindegren et al. 1993b with stacked Fizeau interferometers. A separate telescope, ARVI, for radial velocities has been included as proposed by Favata & Perryman 1995.

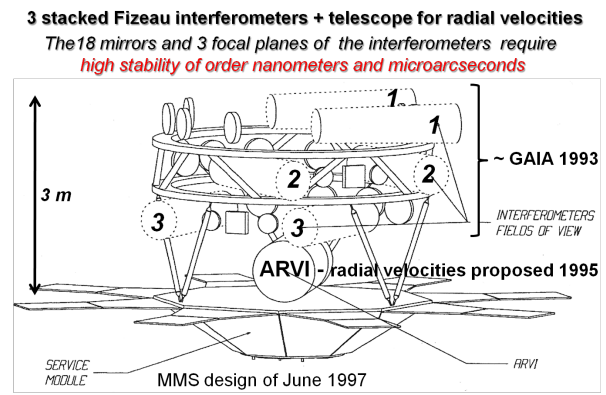

**Figure 18.** The GAIA design of June 1997 as it appeared from the study proposal by MMS (Matra Marconi Space). The detection system with or without a modulating grid, shown in Figures 13 and 14, was dropped by MMS a few months later and by ESA in January 1998 and only detection of the diffraction image directly on the CCD with full telescope aperture was further considered.

It would be interesting to see when the decisions about various important features in the instrument were taken, but it is usually not possible to give a precise time because of the many interrelated issues considered simultaneously. For example, the pros and cons of having three stacked interferometers, as in Figure 18, or only two were discussed at the 4[th] SAG meeting held in Grasse on 24 September 1997, based on a report by Lindegren. Two interferometers remained the baseline.

At the same meeting the requirement of the 90 cm aperture of ARVI was confirmed, with a possible target of 1.3m. MMS was investigating inclusion of ARVI within the interferometer assembly.

I continue to read from the careful minutes by Perryman of this meeting: Technically, the mission target was now a billion objects to a completeness limit of 20 mag. Further studies of the possible photometric system had been made by Høg, with plans for further independent activities to be carried out by Munari (Padova).

Høg presented his latest concept of the sky mapper, sending down information on specific objects. Gilmore presented the case for sending down the entire sky data, which would probably be feasible now with the present sky mapper approach and realistic compression schemes.

Work on on-board detection instead of using an a priori GAIA input catalogue, similar to the Hipparcos input catalogue had already begun in July 1997.

Lattanzi presented the case for a beam combiner (instead of separate interferometers) being studied by Alenia/Aerospatiale/APLT/AMTS.

An outline of the final report, ESA 2000, was presented by Michael Perryman. This is typical for Michael, he was always very timely with preparations for everything, be it international meetings, the team meetings, the subsequent minutes, and in this case with the outline for the final report about the work we had just begun. The outline was in front of us already two years before the report should be completed and this was of course very important for a report which finally contained 381 pages, ESA 2000, the report so crucial for obtaining the mission approval in 2000.

Michael evidently worked all the time on our GAIA project as he had done on Hipparcos since he became leader of that project in 1980. I could always count on him, e.g., to have a long phone conversation after a SAG meeting in order to discuss an issue which had not been adequately solved during the meeting. He could very quickly grasp the essence of any scientific or technical problem. He always arranged a social dinner after the first day of our two-day meetings including engineers from ESA and the industry and I thoroughly enjoyed these dinners with talk of science and many other things. His arrangement of our visit to the Hipparcos launch in South America in 1989 is unforgettable.

The industrial mid-term review was held in ESTEC on 14 January 1998, followed by the 6th SAG meeting of two days. Frédéric Safa, the leading MMS engineer, *"summarised the main lines that the instrumental development had followed over the last two months: (a) the primary was now baselined as a 1.7 x 0.7 m$^2$ monolithic reflector, with an overall f=50m, and resulting in a pixel size along scan of 9µm. (b) the passive telescope design could be achieved without the requirement of nm-accuracy mechanisms. MMS/Safa presented their concepts for the measurement (not control) of basic angle variations at or below the 1 µarcsec level. ..."*

The minutes of the meeting shown in Figure 19, gives the many strong arguments for the monolithic full-aperture reflector, i.e. for abandoning interferometry. Industry, not SAG scientists, had now studied "a large Roemer option", and industry would certainly have found this solution even if we had not proposed the large Roemer to ESA in October 1993. The SAG agreed to the new baseline without anybody thinking of Roemer as far as I know, not even I thought of Roemer. The transition from GAIA with interferometers to a large Roemer can be fixed in time precisely to 15 January 1998.

The last lines in the text of Figure 19 were followed with the inclusion of an interferometric option of Alenia design in the "Red Report", i.e. ESA 2000 - which in fact has a white cover.

---

Extract from the minutes of the SAG meeting on 15 January 1998: *Filling of the central aperture allows the light collecting power to be preserved while decreasing the overall diameter of the payload and spacecraft. Utilization of the 9µm rather than the 6µm pixels allows the payload feasibility to be considerably eased. The monolithic primary allows the structural properties to be enhanced, and leads to a lower data rate (and improved readout noise performance etc) being achieved. The major 'disadvantage' of this approach is that the mission becomes less obviously interferometric. However the central issue is that given the performance figures now achieved with the compact, monolithic, 9µm system, enhanced performance by insisting on an interferometric payload would only come with penalties of detector technology, mass, dimensions, structural stability, and cost, with limited accuracy improvements. At present it is difficult to envisage how such an 'artificial increase' in complexity could be justified. For the final Red Report, these issues could be addressed, with an interferometric option being included for illustration. In parallel, the Alenia design which might be included as an alternative option in the Red Report, is more evidently following more classical interferometric principles.*

**Figure 19.** Arguments for the new GAIA design of January 1998 without interferometric telescopes.

Today, the following reasoning is natural. It is rather trivial that a full-aperture large mirror gives more information of the image position than if the middle part of the large mirror is removed in order to make a Fizeau interferometer. This was put in mathematical form by Lindegren (1998). The author shows: *"... why a single optical aperture is better for a scanning satellite than the originally proposed twin pupil of a Fizeau interferometer."* You cannot improve on astrometry by removing some photons. Secondly, the great complication and consequent high cost of a large interferometer is also obvious considering the requirements on stability of relative positions and angles of many mirrors on the order of nanometers and microarcseconds, cf. Figures 13 and 18. Thirdly, the three stacked interferometers would require an enormous shield against the Sun.

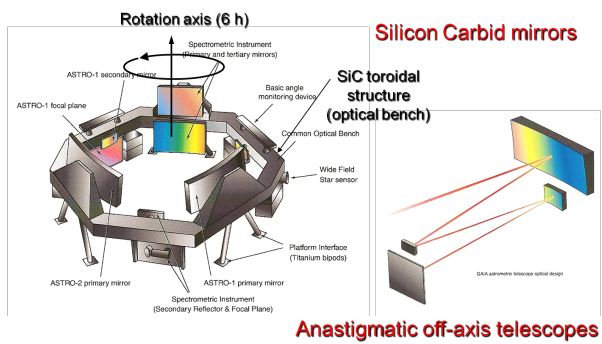

**Figure 20.** The GAIA design of mid 1998. Interferometry has been dropped and the design has become a large Roemer.

These three handicaps make the interferometric GAIA a certain loser against the large Roemer - so it appears with hindsight. But this was not at all clear to me in those years and we never discussed these issues. I worked on the interferometric option, e.g. on optics with GAIA95 and on the detection, Høg 1995c. We all worked with enthusiasm because we believed interferometry could give good astrometry, not just for tactical reasons knowing that interferometry was fashionable in ESA and elsewhere. All of us believed in an interferometric GAIA. We left to the industry to solve the problems we had vaguely seen and the engineers quickly opened our eyes in January 1998.

The design by mid 1998 at the end of the industrial study is shown here in Figure 20. A contemporary status of the GAIA project was presented at the meeting in Gotha in May 1998, available as Lindegren 1998 and Høg et al. 1998. The satellite contains two large telescopes for astrometry instead of three interferometers and a separate telescope of 0.75 x 0.70 $m^2$ aperture for radial velocities and photometry. Photometry in four broad bands is obtained in the astrometric telescopes and in seven medium-width spectral bands in the smaller telescope.

The design is described in great detail in ESA 2000 where the number of intermediate bands has been increased from seven to eleven. The GAIA mission was approved by ESA in 2000 for a launch "not later than 2012." A GAIA Science Team was set up, again led by Michael Perryman. The members up to 2007 were:

| | |
|---|---|
| Frédéric Arenou (2001 - 2005) | Meudon, France |
| Carine Babusiaux (2006 - 2007) | Meudon, France |
| Coryn Bailer-Jones (2001 - 2007) | Heidelberg, Germany |
| Ulrich Bastian (2001 - 2007) | Heidelberg, Germany |
| Anthony Brown (2006 - 2007) | Leiden, The Netherlands |
| Mark Cropper (2006 - 2007) | MSSL - UCL, United Kingdom |
| Erik Høg (2001 - 2007) | Copenhagen, Denmark |
| Andrew Holland (2001 - 2005) | Leicester/Brunel, United Kingdom |
| Carme Jordi (2002-2007) | Barcelona, Spain |
| David Katz (2001 - 2007) | Meudon, France |
| Mario Lattanzi (2001 - 2005) | Torino, Italy |
| Floor van Leeuwen (2003 - 2007) | Cambridge, United Kingdom |
| Lennart Lindegren (2001-2007) | Lund, Sweden |
| Xavier Luri (2001 - 2007) | Barcelona, Spain |
| Francois Mignard (2001-2007) | Nice, France |
| Fred Jansen (2006) | ESA/ESTEC (Gaia Project Scientist) |
| Michael Perryman (2000 - 2006) | ESA/ESTEC (Gaia Project Scientist) |

All the years were busy for my own part with work on many aspects of GAIA, for instance on the design of the optimal photometric system, after 2000 in the chair of the Photometry Working Group together with Carme Jordi. I worked, especially with Frédéric Arenou and Jos de Bruijne, on the optimal sampling or windowing of the CCDs, i.e. the definition of the optimal window of pixels to be transmitted to ground around each star. The windowing means that about 99.9 per cent of the pixels can be skipped since most of the sky contains no stars, even when one billion stars on the sky are detected. This means lower noise in the data and much less data to be transmitted to ground. There are 64 reports since 1997 with authors from Copenhagen about sampling, detection and imaging for GAIA or Gaia.

With collaborators in Copenhagen, C. Fabricius, J. Knude, H.E.P. Lindstrøm, S. Madsen, V.V. Makarov, I.D. Novikov, A.G. Polnarev, H.J. Sørensen, and M. Vaccari, the instrument design and the possibilities to detect and measure certain objects were studied. Signatures of photometric and astrometric microlensing events, supernovae, galaxies, and NEOs were considered, cf references in ESA 2000 and papers in Leiden 1998, Les Houches 2001, and Vilnius 2001. The results were not always promising, but it is important to quantify the possibilities at an early stage where the design might still be adapted. I had learnt this lesson in 1981 when my proposal of the Tycho project, the special use of the Hipparcos star mapper, came almost too late to be implemented in the Hipparcos satellite which had been approved one year before. The numbered reports from Copenhagen on GAIA or Gaia since 1997 reached 185 when my term in the science team ended in 2007, after 32 years in ESA teams on astrometry since October 1975.

The design of optics and detection from 1998 underwent great changes in the years after 2000. The satellite had to be decreased in order to fit inside a Soyuz launcher instead of the Ariane 5 dual launch, and weight and cost problems had to be solved. The 1998 design contained two large and completely separate telescopes for astrometry, each with its own focal plane. The smaller telescope contained two focal planes, one for spectrometry and one for photometry. Several different CCDs with different size of pixels were required.

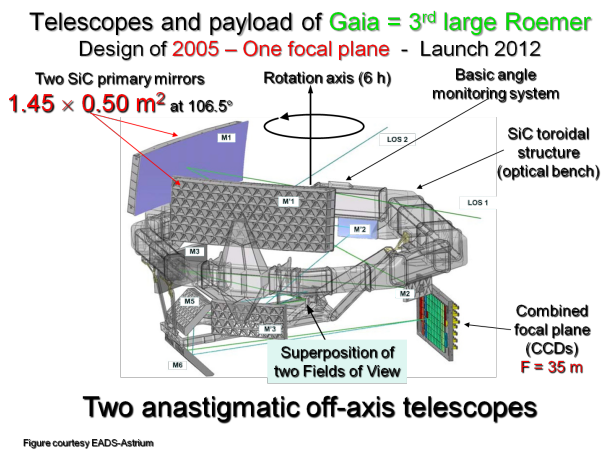

**Figure 21.** The Gaia design of 2005 by EADS-Astrium.

The final design of 2005 is very much simpler and fitting the ESA cost envelope, which is in fact the same as for the Hipparcos mission when transferred to the same economic situation. There are only two telescopes of 1.45 x 0.50 m$^2$ apertures and only one focal plane of 0.7 x 0.7 degrees holding 106 large-format CCDs, performing: star detection, astrometry, photometry and spectrometry. Only one type of CCD though with different sensitization is required. The beams from the two primary mirrors are separated by a 'basic angle' of 106.5 degrees and are brought together by a beam combiner placed at the intermediate focus. The filter photometry has been replaced by low-dispersion spectrophotometry. The penalty for astrometry has been an increased expected standard error at, e.g., V=15 mag from 11 μarcsec to about 25 which is still within the goal set originally.

## 9. The future

The present report describes the chain of ideas and actions which lead to the proposals of Roemer, GAIA and Gaia in the 1990s, especially to the evolution of the various optics and detector systems. The scientific and technical environment has been included so that the development can be properly understood. It must not be read as a history of the Gaia project, nor of course as a history of space astrometry in total, but such histories should be written. - I will end here with a view of the future.

The Gaia mission will deliver astrometric data of high accuracy, beginning a few years after launch and with final data by 2020. The astrometry will be global, covering the entire sky to 20$^{th}$ magnitude with stellar distances, positions and proper motions for astrophysical and all kinds of use. The photometric data for the same one billion stars at 100 epochs during the five or six year mission will provide a unique census for study of stellar variability. This data set will be unrivalled in its kind for several decades since it is difficult to imagine that any space agency will approve a new and better mission and be ready to launch before 2040, considering for instance the great difficulties encountered at the approval of Hipparcos in 1980 and of the Cornerstone study in 1997.

On such time scales, it is hard to imagine that the astrometric expertise in the present Gaia teams can be preserved for a new mission. This was much easier for Gaia because a realistic design of a new mission was available already in 1992 while Hipparcos was still operating. This was possible since CCD detectors were highly developed at that time and it could be seen that they would be much more efficient in an astrometric satellite than the photoelectric detectors used in Hipparcos. No similar technological basis for a great improvement is known to me, but a new mission similar to Gaia should be considered.

All-sky scanning satellites as Hipparcos and Gaia cannot stop and stare at selected stars or areas. This requires pointing satellites and such missions for limited sky coverage and a smaller number of stars, but with higher accuracy and/or at redder wavelengths may have a better chance of approval.

Higher accuracy can in principle be obtained by more photons (i.e. longer integration times and/or larger apertures) and/or higher angular resolution, for instance by interferometry. It is difficult to imagine that NASA or any other space agency will soon engage in interferometry for astrometry after the long effort on SIM has been stopped, SIM 2011, and after the lesson from the study of the interferometric version of GAIA 1993-97.

The long-term Japanese plans for high-accuracy infrared astrometry with Jasmine of the Galactic bulge look promising, Gouda et al. 2009 and Jasmine 2011. A small low-cost scanning satellite called "Nano-JASMINE" is due for launch in August 2011. A "Small Jasmine" and "Jasmine" are both pointed satellites and launches are expected in 2016 and in the 2020s.

The J-MAPS astrometric mission by the USNO is a pointing satellite using hybrid CMOS/CCD detectors, but it is an all sky mission using the overlapping plate solution method. It is expected to be launched in 2014 and it will observe stars to 14$^{th}$ magnitude, Hennessy & Gaume 2009.

**Acknowledgements:** I would like to thank Timo Prusti and his collaborators for the Gaia schedule *without Roemer* and Francois Mignard for his slide with Roemer as one of the *"unsuccessful followers"* of Hipparcos because they thereby got me started on the present report, and I thank Francois and Timo for their warm welcome at my subsequent lectures on this subject at respectively CNES in Toulouse and at ESTEC. I am grateful to Ulrich Bastian, Lars Brink, Anthony Brown, Tom Corbin, Aase Høg, Carme Jordi, Lennart Lindegren, Francois Mignard, Michael Perryman, Rudolf Le Poole, Ken Seidelmann, and Sperello di Serego Alighieri for correspondence during the recent months about the subject and for comments to previous versions of this report.

## 10. References on space astrometry

The following list was compiled on the development of space astrometry especially in the 1990s, but more references are included than mentioned in the present report *though without being complete.*

The references are given in the sequence they were presented publicly at conferences or by other distribution. The year of publication in proceedings is often a year later. Reports on Hipparcos and Tycho at the symposia are mostly omitted.

**1980:**
Di Serego Alighieri S., Rodgers A.W. & Stapinski T.E. 1980, An alternative detector for Hipparcos. Padova – Asiago Obs. / M.S.S.S.O. Technical Report No. 2, January 1980. As received from M. Perryman in 2011, available at www.astro.ku.dk/~erik/AltDetHip.pdf

**Leningrad in October 1989:**
Chubey M.S., Makarov V.V., Yershov V.N. Kanayev I.I., Fomin V.A., Streletsky Yu.S., Schumacher A.V. 1990, A Proposal for Astrometric Observations from Space. In: J.H. Lieske and V.K. Abalakin (eds.) *Inertial Coordinate Systems on the Sky.* Leningrad, U.S.S.R., Oct. 17-21, 1989, IAU Symp. No. 141, 77.

Chandler J.F. and Reasenberg R.D. 1990, **POINTS:** A Global Reference Frame Opportunity. In: J.H. Lieske and V.K. Abalakin (eds.) *Inertial Coordinate Systems on the Sky.* Leningrad, U.S.S.R., Oct. 17-21, 1989, IAU Symp. No. 141, 217.

Duncombe R.L., Jefferys W.H., Benedict G.F., Hemenway P.D., Shelus P.J.. 1990, Expectations for Astrometry with the Hubble Space Telescope. In: J.H. Lieske and V.K. Abalakin (eds.) *Inertial Coordinate Systems on the Sky.* Leningrad, U.S.S.R., Oct. 17-21, 1989, IAU Symp. No. 141, 339.

Seidelmann P.K. 1990, Space Astrometry and the HST Wide Field/Planetary Camera. In: J.H. Lieske and V.K. Abalakin (eds.) *Inertial Coordinate Systems on the Sky.* Leningrad, U.S.S.R., Oct. 17-21, 1989, IAU Symp. No. 141, 347.

Nesterov V.V., Ovchinnikov A.A., Cherespachuk A.M., Sheffer E.K. 1990, The **LOMONOSOV** Project for Space Astrometry. In: J.H. Lieske and V.K. Abalakin (eds.) *Inertial Coordinate Systems on the Sky.* Leningrad, U.S.S.R., Oct. 17-21, 1989, IAU Symp. No. 141, 355.

Avanesov G.A., Vavaev V.A., Ziman Ya.L et al. 1990, **REGATTA-ASTRO** Project: Astrometric studies from Small Space Laboratory. In: J.H. Lieske and V.K. Abalakin (eds.) *Inertial Coordinate Systems on the Sky.* Leningrad, U.S.S.R., Oct. 17-21, 1989, IAU Symp. No. 141, 361.

Stone R.C. and Monet D.G. 1990, The USNO (Flagstaff Station) CCD Transit Telescope and Star Positions Measured from Extragalactic Sources. In: J.H. Lieske and V.K. Abalakin (eds.) *Inertial Coordinate Systems on the Sky.* Leningrad, U.S.S.R., Oct. 17-21, 1989, IAU Symp. No. 141, 369.

**Moscow in June 1991:**
Høg E. & Chubey M. 1991. Proposal for a second Hipparcos. Presented at the International Symposium "Etalon" Satellites held in June 1991 in Moscow and as poster at the IAU General Assembly in Buenos Aires in August. It was not published but is now scanned and the 6 pages are available at www.astro.ku.dk/~erik/Hipparcos-2.pdf

**Shanghai in September 1992:**
Høg E. and Lindegren L. 1993, A CCD Modulation Detector for a Second Hipparcos Mission. In: I.I. Mueller and B. Kolaczek (eds.) *Developments in Astrometry and Their Impact on Astrophysics and Geodynamics.* Shanghai, China, Sept. 15-19, 1992, IAU Symp. No. 156, 31. At http://esoads.eso.org/abs/1993IAUS..156...31H

Høg E. 1993, Astrometry and Photometry of 400 Million Stars Brighter than 18 Mag. In: I.I. Mueller and B. Kolaczek (eds.) *Developments in Astrometry and Their Impact on Astrophysics and Geodynamics.* Shanghai, China, Sept. 15-19, 1992, IAU Symp. No. 156, 37. Available at http://esoads.eso.org/abs/1993IAUS..156...37H

**Proposal for ESA M3 mission in May 1993:**
Lindegren L. (ed.), Bastian U., Gilmore G., Halbwachs J.L., Høg E., Knude J., Kovalevsky J., Labeyrie A., van Leeuwen F., Pel J.W., Schrijver H., Stabell R., Thejll P. 1993a. ROEMER: Proposal for the Third Medium Size ESA Mission (M3). Lund Observatory.

**Cambridge in June 1993:**
Morrison and G.F. Gilmore (eds.) 1994, *Galactic and Solar System Optical Astrometry.* Proceedings of a conference held in Cambridge June 21-24, 1993. 19+339 pp.

Høg E. and Lindegren L. 1994, ROEMER satellite project: the first high-accuracy survey of faint stars. In: L.V.

Morrison and G.F. Gilmore (eds.) *Galactic and Solar System Optical Astrometry*. Proceedings of a conference held in Cambridge June 21-24, 1993. 246-252.

**To ESA in October 1993:**
Lindegren L., Perryman M.A.C., Bastian U., Dainty J.C., Høg E., van Leeuwen F., Kovalevsky J., Labeyrie A., Mignard F., Noordam J.E., Le Poole R.S., Thejll P., Vakili F. 1993b. GAIA: Global Astrometric Interferometer for Astrophysics, proposal for a Cornerstone Mission concept submitted to ESA on 12 October 1993, including a cover letter by Lindegren, 6 pages. Available at www.astro.ku.dk/~erik/gaia_proposal_1993.pdf

**March 10, 1994:**
Seidelmann writes to me that USNO is pursuing NEWCOMB which is however unfunded, and that they are interested in collaboration on other space-based astrometry projects like Roemer and GAIA.

**April 1994:**
Lindegren L. 1994. Letter in April 1994 to Steven Beckwith, chairman of the UV-to-radio topical team of the Survey Committee then drafting ESA's Horizon 2000+ plan, 1+7 pages. Available at www.astro.ku.dk/~erik/beckwith_lindegren.pdf

**The Hague in August 1994:**
Høg E. & Seidelmann P.K. (eds.) 1995, *Astronomical and Astrophysical Objectives of Sub-milliarcsecond Optical Astrometry,* Proceedings of the 166th Symposium of the IAU, held in The Hague, The Netherlands, August 15-19, 1994,16+441 pp.

Høg E. 1995a, A New Era of Global Astrometry. II: A 10 Microarcsecond Mission. (Including proposal of Roemer+). In: Proceedings of the 166th Symposium of the IAU, 317.

Johnston K., Seidelmann P.K., Reasenberg R.D., Babcock R., Philips J.D. 1995, Newcomb Astrometric Satellite. In: Proceedings of the 166th Symposium of the IAU, 331.

Lindegren L. and Perryman M.A.C. 1995, A Small Interferometer in Space for Global Astrometry: The GAIA Concept. In: Proc. of the 166th Symposium of the IAU, 337.

**To ESA in September 1994:**
Lindegren L. & Perryman M.A.C. 1994. GAIA: Global Astrometric Interferometer for Astrophysics (A Concept for an ESA Cornerstone Mission). Supplementary Information Submitted to the Horizon 2000+ Survey Committee.

**Cambridge in June 1995:**
Perryman M.A.C. & van Leeuwen F., eds. 1995, Proceedings of a Joint RGO-ESO Workshop on "Future Possibilities for Astrometry in Space", Cambridge, UK, 19-21 June 1995 (**ESA SP-379**, September 1995), 323 pp.

Some of the following reports are available at www.astro.ku.dk/~erik/papers/

Cecconi M. & Lattanzi M.G. 1995, GAIA Optics: Polychromatic PSF and Pupils. In: ESA SP-379, September 1995, 251.

Daigne G. 1995, Direct Fringe Detection and Sampling of the Diffraction Pattern. In: ESA SP-379, September 1995, 209.

Fabricius C. & Høg E. 1995, Scientific Analysis of Data from a Scanning Satellite. In: ESA SP-379, September 1995, 273.

Favata F. & Perryman M. 1995, Parallel Aquisition of Radial Velocities and Metallicities for a GAIA-Type Mission. In: (ESA SP-379, September 1995, 153.

Gai M., Lattanzi M.G., Casertano S. & Guarnieri M.D. 1995, Non-Conventional Detector Applications for Direct Focal Plane Coverage. In: ESA SP-379, September 1995, 231.

Gilmore G. & Høg E. 1995, Key Questions in Galactic Structure with Astrometric answers. In: ESA SP-379, September 1995, 95.

Høg E. 1995b, Observation of Microlensing by an Astrometric Satellite. In: ESA SP-379, September 1995, 125.

Høg E. 1995c, Some Designs of the GAIA Detector System. In: ESA SP-379, September 1995, 223.

Høg E. 1995d, Three Astrometric Mission Options and a Photometric System. In: ESA SP-379, September 1995, 255.

Høg E. 1995e, Astrometric Satellite with Sunshield. In: ESA SP-379, September 1995, 263.

Lindegren L. 1995, Summary of the Parallel Sessions. In: ESA SP-379, September 1995, 267.

Lindegren L. & Perryman M.A.C. 1995, The GAIA Concept. In: ESA SP-379, September 1995, 23.

Loiseau S. & Shaklan S. 1995, Analysis of an Astrometric Fizeau Interferometer for GAIA. In: ESA SP-379, September 1995, 241.

Makarov V. 1995c, Gravitational Waves as Astrometric Targets. In: ESA SP-379, September 1995, 117.

Noordam J.E. 1995, On the Advantage of Dispersed Fringes. In: ESA SP-379, September 1995, 213.

Perryman M.A.C. & Peacock A. 1995,A Superconducting Detector for a Future Space Astrometry Mission. In: ESA SP-379, September 1995, 207.

Rabbia Y. 1995, Obtaining Spectral Information from an Astrometric Interferometric Mission. In: ESA SP-379, September 1995, 217.

Seidelmann P.K. et al. 1995, A Fizeau Optical Interferometer Astrometric Satellite. In: ESA SP-379, September 1995, 187.

Straizys V. & Høg E. 1995, An Optimum Eight-Colour Photometric System for a Survey Satellite. In: ESA SP-379, September 1995, 191.

Yershov V.N. 1995, An Ultraviolet Option for a Future Astrometric Satellite. In: ESA SP-379, September 1995, 197.

**FAME-1 in 1995:**
Johnston K. 1995a June, Step 1 Proposal to NASA for the Medium-class Explorer (MIDEX), Fizeau Astrometric Mapping Explorer (FAME). Technical Report with 40 pages.

Johnston K. 1995b December, Step 2 Proposal to NASA for the Medium-class Explorer (MIDEX), Fizeau Astrometric Mapping Explorer (FAME). Technical Report with 25 pages Technical Volume.

**GAIA95 report in November 1995:**
Høg E., Fabricius C., Makarov V.V. 1995, GAIA95: A New Optical Design for the GAIA Astrometry Mission. A report from Copenhagen. A nearly identical draft from 3 November was called "Mapping Interferometer for Astrometry and Spectrophotometry", both versions were widely distributed. It appeared in Experimental Astronomy in 1997.

Høg E., Fabricius C., Makarov V.V. 1997, GAIA95: Astrometry from Space: New Design of the GAIA Mission. Experimental Astronomy 7: 101.

**1996:**
Röser S., Bastian U., de Boer K.S., Høg E., Schalinski C., Schilbach E., de Vegt Ch., Wagner S. 1996, August. DIVA: Deutsches Interferometer für Vielkanalphotometrie und Astrometrie. Antrag für die Phase A Studie.

Seidelmann P.K. 1998, Prospects for Future Astrometric Missions. In: B.J. McLean et al. (eds.), *New Horizons from Multi-Wavelength Sky Surveys,* Proceedings of the 179th Symposium of the IAU, held in Baltimore, U.S.A, August 26-30, 1996, 79.

**1997:**
Cecconi M., Gai M. & Lattanzi M.G. 1997, A New Optical Configuration for GAIA. In: ESA SP-402, 803.

Cecconi M., Rigoni G. & Bernacca P.L. 1997, Opto/Mechanical Analysis for the Space Interferometry Project GAIA. In: ESA SP-402, 811.

ESA 1997, Invitation to join the ESA Science Advisory Group on Astrometry Cornerstone Study. Available at www.astro.ku.dk/~erik/esa_sag_invitation_970311.pdf

Gai M., Bertinetto F., Bisi M., et al. 1997, GAIA Feasibility: Current Research on Critical Aspects. In: ESA SP-402, 835.

Høg E., Bastian U. & Seifert W. 1997, Optical Design for GAIA. In: ESA SP-402, 783.

Lindegren L. & Perryman M.A.C. 1997, GAIA: Global Astrometric Interferometer for Astrophysics. In: ESA SP-402, 799.

Perryman M.A.C. & Bernacca P.L. (eds.) 1997, HIPPARCOS Venice '97, Presentation of the Hipparcos and Tycho Catalogues. ESA SP-402, 52+862 pp.

Reasenberg R.D., Babcock R.W., Chandler J.F., Phillips J.D. 1997, POINTS Mission Studies: Lessons for SIM. Center for Astrophysics Preprint Series No. 4496 (Received February 28, 1997). Invited paper for a conference at STEcI in October 1996. 15 pages. It contains e.g. some instrument descriptions of POINTS and NEWCOMB but no designs.

Scholz R.-D. & Bastian U. 1997, Simulated Dispersed Fringes of an Astrometric Space Interferometry Mission. In: ESA SP-402, 815.

**Gotha in May 1998:**
Lindegren L. 1998, Hipparcos and the future: GAIA. In: P. Brosche et al. (eds) Proceedings of the International Spring Meeting of the Astronomische Gesellschaft, Gotha, May 11-15, 1998, "The Message of the Angles – Astrometry from 1798 to 1998", Acta Historica Astronomiae, vol. 3, 214. http://esoads.eso.org/abs/1998AcHA....3..214L

Høg E., Fabricius C., Knude J., Makarov V.V. 1998, Sky survey and photometry by the GAIA satellite. In: P. Brosche et al. (eds) Proceedings of the International Spring Meeting of the Astronomische Gesellschaft, Gotha, May 11-15, 1998, "The Message of the Angles – Astrometry from 1798 to 1998", Acta Historica Astronomiae, vol. 3, 223. http://esoads.eso.org/abs/1998AcHA....3..223H

**Leiden in November 1998:**
Leiden 1998, Proceedings of the GAIA Workshop, November 23-27, 1998 in Leiden. In: Baltic Astronomy, An international journal, Vol.8, Nrs.1 and 2, 324 pp, 1999.

**2000:**
ESA 2000, GAIA, Composition, Formation and Evolution of the Galaxy, Concept and Technology Study Report. ESA-SCI(2000)4, 381pp. Available in "Library &Livelink" at www.rssd.esa.int/index.php?project=GAIA&page=index

**Les Houches in May 2001:**
Les Houches 2001, GAIA: A European Space Project – A summer school 14-18 May 2001. O. Bienayme & C. Turon (eds.), EAS Publication Series, Volume 2, 395 pp, 2002.

**Vilnius in July 2001:**
Vilnius 2001, Census of the Galaxy: Challenges for Photometry and Spectrometry with GAIA. Vansevicius V., Kucinskas A., Sudzius J. (eds.), Proceedings of the Workshop held in Vilnius, Lithuania, 2-6 July 2001. From: Astrophysics and Space Sciences Vol. 280, Nos. 1-2, 194 pp, 2002.

**2003:**
Johnston K.J. 2003, The FAME Mission. In: Proceedings of the SPIE, Volume 4854, 303. At 2003SPIE.4854..303J

**2007:**
Høg E. 2007, From the Roemer mission to Gaia. A poster at IAU Symposium No. 248 in Shanghai, October 2007. Only the first three pages appear in the Proceedings. Complete version available at www.astro.ku.dk/~erik/ShanghaiPoster.pdf and as report No.4 at www.astro.ku.dk/~erik/History.pdf

**2008:**
Høg E. 2008a, Bengt Strömgren and modern astronomy: Development of photoelectric astrometry including the Hipparcos mission. At www.astro.ku.dk/~erik/History.pdf and as poster at IAU Symposium No. 254.

Høg E. 2008b, Astrometric accuracy during 2000 years. www.astro.ku.dk/~erik/Accuracy.pdf

Høg E. 2008c, Astrometry and Optics during the Past 2000 Years, a collection of 9 reports. At www.astro.ku.dk/~erik/History.pdf.

Høg E. 2008d, Four lectures about the General History of Astrometry. At www.astro.ku.dk/~erik/Lectures.pdf

**2009:**
Gouda N. et al. 2009, Series of JASMINE projects ---
Exploration of the Galactic bulge ---
http://www.ast.cam.ac.uk/iau_comm8/iau27/presentations/Gouda.pdf

Hennessy G.S. & Gaume R. 2009, Space astrometry with the Joint Milliarcsecond Astrometry Pathfinder. Proc. of the IAU Symposium 261, 350.

**2011:**
Høg E. 2011a, Miraculous approval of Hipparcos in 1980: (2). At www.astro.ku.dk/~erik/HipApproval.pdf

Høg E. 2011b, Astrometry Lost and Regained. Or: From a modest experiment in Copenhagen in 1925 to the Hipparcos and Gaia space missions. At
www.astro.ku.dk/~erik/AstromRega3.pdf

Høg E. 2011c, Lectures on Astrometry.
www.astro.ku.dk/~erik/Lectures2.pdf

Jasmine 2011, European-Japanese collaboration. Nano-JASMINE and AGIS
www.rssd.esa.int/index.php?project=GAIA&page=picture_of_the_week&pow=132

SIM 2011,
http://en.wikipedia.org/wiki/Space_Interferometry_Mission

# Astrometry and optics during the past 2000 years

Erik Høg     Niels Bohr Institute, Copenhagen, Denmark

2011.05.03:   Collection of reports from November 2008

ABSTRACT: The satellite missions Hipparcos and Gaia by the European Space Agency will together bring a decrease of astrometric errors by a factor 10000, four orders of magnitude, more than was achieved during the preceding 500 years. This modern development of astrometry was at first obtained by photoelectric astrometry. An experiment with this technique in 1925 led to the Hipparcos satellite mission in the years 1989-93 as described in the following reports Nos. 1 and 10. The report No. 11 is about the subsequent period of space astrometry with CCDs in a scanning satellite. This period began in 1992 with my proposal of a mission called Roemer, which led to the Gaia mission due for launch in 2013. My contributions to the history of astrometry and optics are based on 50 years of work in the field of astrometry but the reports cover spans of time within the past 2000 years, e.g., 400 years of astrometry, 650 years of optics, and the "miraculous" approval of the Hipparcos satellite mission during a few months of 1980.

2011.05.03:   Collection of reports from November 2008. The following contains overview and link to the reports Nos. 1-9 from 2008 and Nos. 10-13 from 2011.
Direct quick connection to individual reports: www.astro.ku.dk/~erik/1104-index.pdf .
The reports are collected in two big file, see details on p.8.

# CONTENTS of Nos. 1-9 from 2008

# CONTENTS of Nos. 10-13 from 2011

# Overview with links to Nos. 1-9

No. 1 -  2008.05.27:

## Bengt Strömgren and modern astrometry:
## Development of photoelectric astrometry
## including the Hipparcos mission

ABSTRACT:  Bengt Strömgren is known as the famous astrophysicist and as a leading figure in many astronomical enterprises. Less well-known, perhaps, is his role in modern astrometry although this is equally significant. There is an unbroken chain of actions from his ideas and experiments with photoelectric astrometry since 1925 over the new meridian circle in Denmark in the 1950s up to the Hipparcos and Tycho Catalogues published in 1997.
www.astro.ku.dk/~erik/Stroemgren.pdf
Contribution to IAU Symposium No. 254 in Copenhagen, June 2008: The Galaxy Disk in Cosmological Context – Dedicated to Professor Bengt Strömgren (1908-1987).

No. 1A - 2008.06.10:

## Bengt Strömgren and modern astrometry ... (Short version)

www.astro.ku.dk/~erik/StroemgrenShort.pdf
The same title as No. 1, but containing the short version posted at the symposium.

No. 2 - 2008.03.31:

## Lennart Lindegren's first years with Hipparcos

ABSTRACT: Lennart Lindegren has played a crucial role in the Hipparcos project ever since he entered the scene of space astrometry in September 1976. This is an account of what I saw during Lennart's first years in astrometry after I met him in 1976  when he was a young student in Lund.
www.astro.ku.dk/~erik/Lindegren.pdf

No. 3 – 2008.05.28:

## Miraculous approval of Hipparcos in 1980

ABSTRACT: The approval of the Hipparcos mission in 1980 was far from being smooth since very serious hurdles were encountered in the ESA committees. This process is illuminated here by means of documents from the time and by recent correspondence. The evidence leads to conclude that in case the approval would have failed, Hipparcos or a similar scanning astrometry mission would never have been realized, neither in Europe nor anywhere else.
www.astro.ku.dk/~erik/HipApproval.pdf

No. 4 -  2007.12.10:

## From the Roemer mission to Gaia

ABSTRACT: At the astrometry symposium in Shanghai 1992 the present author made the first proposal for a specific mission concept post-Hipparcos, the first scanning astrometry mission with CCDs in time-delayed integration mode (TDI). Direct imaging on CCDs in long-focus telescopes was described as later adopted for the Gaia mission. The mission called Roemer was designed to provide accurate astrometry and multi-colour photometry of 400 million stars brighter than 18 mag in a five-year mission. The early years of this mission concept are reviewed.
www.astro.ku.dk/~erik/ShanghaiPoster.pdf
Presented as poster at IAU Symposium No. 248 in Shanghai, October 2007. Only the first three pages appear in the Proceedings.

No. 5 -  2008.05.23, updated 2008.11.25.
Note in 2011: See further update in **www.astro.ku.dk/~erik/History2**

# Four lectures on the general history of astrometry

Overview, handout, abstracts at:   www.astro.ku.dk/~erik/Lectures.pdf
**Brief overview :**
 Lecture No. 1:

### Astrometry and photometry from space: Hipparcos, Tycho, Gaia

  The introduction covers 2000 years of astronomy from Ptolemy to modern times. The Hipparcos mission of the European Space Agency was launched in 1989, including the Tycho experiment. The Hipparcos mission and the even more powerful Gaia mission to be launched in 2011 are described.

 Lecture No. 2:

### From punched cards to satellites: Hipparcos, Tycho, Gaia

  A personal review of 54 years development of astrometry in which I participated.

 Lecture No. 3:

### The Depth of Heavens - Belief and Knowledge during 2500 Years

  The lecture outlines the understanding of the structure of the universe and the development of science during 5000 years, focusing on the concept of distances in the universe and its dramatic change in the developing cultural environment from Babylon and ancient Greece to modern Europe.

 Lecture No. 4, included on 2008.11.25:

### 400 Years of Astrometry: From Tycho Brahe to Hipparcos

  Four centuries of techniques and results are reviewed, from the pre-telescopic era up to the use of photoelectric astrometry and space technology in the first astrometric satellite, Hipparcos, launched by ESA in 1989. The lecture was presented as invited contribution to the symposium at ESTEC in September 2008: **400 Years of Astronomical Telescopes: A Review of History, Science and Technology.** The report submitted to the proceedings is included as No. 8 among "Contributions to the history of astrometry".

No. 6 – 2008.11.25:

## Selected astrometric catalogues

ABSTRACT: A selection of astrometric catalogues are presented in three tables for respectively positions, proper motions and trigonometric parallaxes. The tables contain characteristics of each catalogue to show especially the evolution over the last 400 years in optical astrometry. The number of stars and the accuracy are summarized by the weight of a catalogue, proportional with the number of stars and the statistical weight.
www.astro.ku.dk/~erik/AstrometricCats.pdf

No. 7 – 2008.11.25:

## Astrometric accuracy during the past 2000 years

ABSTRACT: The development of astrometric accuracy since the observations by Hipparchus, about 150 B.C., has been tremendous and the evolution has often been displayed in a diagram of accuracy versus time. Some of these diagrams are shown and the quite significant differences are discussed. A new diagram is recommended and documented.
www.astro.ku.dk/~erik/Accuracy.pdf
The two diagrams, Fig. 1a and 1b, in black/white and colour :
www.astro.ku.dk/~erik/AccurBasic.pdf    www.astro.ku.dk/~erik/AccuracyColour.jpg
www.astro.ku.dk/~erik/AccuracyBW.wmf    www.astro.ku.dk/~erik/AccuracyColour.wmf

No. 8 -  2008.11.25:

## 400 Years of Astrometry: From Tycho Brahe to Hipparcos

ABSTRACT: Galileo Galilei's use of the newly invented telescope for astronomical observation resulted immediately in epochal discoveries about the physical nature of celestial bodies, but the advantage for astrometry came much later. The quadrant and sextant were pre-telescopic instruments for measurement of large angles between stars, improved by Tycho Brahe in the years 1570-1590. Fitted with telescopic sights after 1660, such instruments were quite successful, especially in the hands of John Flamsteed. The meridian circle was a new type of astrometric instrument, already invented and used by Ole Rømer in about 1705, but it took a hundred years before it could fully take over. The centuries-long evolution of techniques is reviewed, including the use of photoelectric astrometry and space technology in the first astrometry satellite, Hipparcos, launched by ESA in 1989. Hipparcos made accurate measurement of large angles a million times more efficiently than could be done in about 1950 from the ground, and it will soon be followed by Gaia which is expected to be another one million times more efficient for optical astrometry.
www.astro.ku.dk/~erik/Astrometry400.pdf
Invited contribution to the symposium in Leiden in October 2008:
**400 Years of Astronomical Telescopes: A Review of History, Science and Technology**

No. 9 -  2008.11.25:

## 650 Years of Optics: From Alhazen to Fermat and Rømer

ABSTRACT: Under house arrest in Cairo from 1010 to 1021, Alhazen wrote his Book of Optics in seven volumes. (The caliph al-Hakim had condemned him for madness.) Some parts of the book came to Europe about 1200, were translated into Latin, and had great impact on the development of European science in the following centuries. Alhazen's book was considered the most important book on optics until Johannes Kepler's "Astronomiae Pars Optica" in 1604. Alhazen's idea about a finite speed of light led to "Fermat's principle" in 1657, the foundation of geometrical optics.
www.astro.ku.dk/~erik/HoegAlhazen.pdf
Contribution to the symposium in Leiden in September 2008:
**400 Years of Astronomical Telescopes: A Review of History, Science and Technology**

# Overview with links to Nos. 10-13

No. 3.2 – 2011.01.27, update from a version of 2008.05.27:

## Miraculous approval of Hipparcos in 1980: (2)

ABSTRACT: The approval of the Hipparcos mission in 1980 was far from being smooth since very serious hurdles were encountered in the ESA committees. This process is illuminated here by means of documents from the time and by recent correspondence. The evidence leads to conclude that in case the approval would have failed, Hipparcos or a similar scanning astrometry mission would never have been realized, neither in Europe nor anywhere else.
www.astro.ku.dk/~erik/HipApproval.pdf

No. 10 - 2011.03.26:

# Astrometry Lost and Regained

## From a modest experiment in Copenhagen in 1925 to the Hipparcos and Gaia space missions

ABSTRACT: Technological and scientific developments during the past century made a new branch of astronomy flourish, i.e. astrophysics, and resulted in our present deep understanding of the whole Universe. But this brought astrometry almost to extinction because it was considered to be dull and old-fashioned, especially by young astronomers. Astrometry is the much older branch of astronomy which performs accurate measurements of positions, motions and distances of stars and other celestial bodies. Astrometric data are of great scientific and practical importance for investigation of celestial phenomena and also for control of telescopes and satellites and for monitoring of Earth rotation. Our main subject is the development during the 20$^{th}$ century which finally made astrometry flourish as an integral part of astronomy through the success of the Hipparcos astrometric satellite, soon to be followed by the even more powerful Gaia mission.
www.astro.ku.dk/~erik/AstromRega3.pdf

No. 11 - 2011.04.06:

# Roemer and Gaia

ABSTRACT: During the Hipparcos mission in September 1992, I presented a concept for using direct imaging on CCDs in scanning mode in a new and very powerful astrometric satellite, Roemer. The Roemer concept with larger aperture telescopes for higher accuracy was developed by ESA and a mission was approved in 2000, expected to be a million times better than Hipparcos. The present name Gaia for the mission reminds of an interferometric option also studied in the period 1993-97, and the evolution of optics and detection in this period is the main subject of the present report. The transition from an interferometric GAIA to a large Roemer was made on 15 January 1998. It will be shown that without the interferometric GAIA option, ESA would hardly have selected astrometry for a Cornerstone study in 1997, and consequently we would not have had the Roemer/Gaia mission.
www.astro.ku.dk/~erik/RoemerGaia.pdf

No. 12 - 2011.01.15:   On the website of the Niels Bohr Institute:

# Surveying the sky

"An astrometric experiment in 1925 was the beginning of a development which Erik Høg, Associate Professor Emeritus, took part in for 50 years. A scientific highlight is the star catalogue Tycho-2 from the year 2000, which describes the positions and movements of 2.5 million stars and is now absolutely essential to controlling satellites and for astronomical observations."

In English:  http://www.nbi.ku.dk/english/www/   and   in Danish:  http://www.nbi.ku.dk/hhh/

and

# En landmåler i himlen

In Danish: En artikel i tidsskriftet KVANT, oktober 2010, om 50 års arbejde

Erindringer om 50 år med astrometrien, der begyndte ved en høstak syd for Holbæk og førte til bygning af to satellitter. Et videnskabeligt højdepunkt er stjernekataloget Tycho-2, der nu er helt uundværligt ved styring af satellitter og ved astronomiske observationer.
www.astro.ku.dk/~erik/kv-2010-3-EH-astrometri.pdf

No. 13 - 2011.03.26:

# Lectures on astrometry

Overview, handout, abstracts at:   www.astro.ku.dk/~erik/Lectures2.pdf
## Brief overview :

**Lecture No. 1.**  45 minutes
 **Astrometry Lost and Regained**
   **From a modest experiment in Copenhagen in 1925**
   **to the Hipparcos and Gaia space missions**
  The lecture has been developed over many years and was held in, e.g., Copenhagen, Vienna, Bonn, Düsseldorf, Vilnius, Oslo, Nikolaev, Poltava, Kiev, Thessaloniki, Ioannina, Athens, Rome, Madrid, Washington, and Charlottesville - since 2007 in PowerPoint.  Revised in 2009 and with the new title

*Astrometry Lost and Regained* it was held in Heidelberg, Sct. Petersburg, Rio de Janeiro, Morelia, Mexico City, Beijing, Montpellier, Groningen, Amsterdam, and Leiden.

---

**Lecture No. 2.** 45 minutes
**Hipparcos - Roemer - Gaia**
 The lectures briefly outlines the development of photoelectric astrometry culminating with the Hipparcos mission. Development of the Gaia mission beginning in 1992 is followed in detail.
 The lecture has been held since 2010 in Toulouse and at ESTEC in Holland.
______________________________________________________________

**Lecture No. 3.** 45 minutes. Suited for a broad audience, including non-astronomers
**The Depth of Heavens - Belief and Knowledge during 2500 Years**
 The lecture outlines the structure of the universe and the development of science during 5000 years, focusing on the distances in the universe and their dramatic change in the developing cultural environment from Babylon and ancient Greece to modern Europe.
 The lecture was first held in 2002, and since 2007 in PowerPoint. Held in Copenhagen, Vilnius, Nikolaev, Athens, Catania, Madrid, and Paris
  Handouts at: www.astro.ku.dk/~erik/DepthHeavens2.pdf
  and  www.astro.ku.dk/~erik/DepthHeavens.pdf

 **An article with the same title as the lecture** appeared in Europhysics News (2004) Vol. 35 No.3. Here slightly updated, 2004.02.20:  www.astro.ku.dk/~erik/Univ7.5.pdf

---

**Lecture No. 4.** 45 or 30 minutes.
**400 Years of Astrometry: From Tycho Brahe to Hipparcos**
 The four centuries of techniques and results are reviewed, from the pre-telescopic era until the use of photoelectric astrometry and space technology in the first astrometry satellite, Hipparcos, launched by ESA in 1989.
 The lecture was presented as invited contribution to the symposium at ESTEC in September 2008: **400 Years of Astronomical Telescopes: A Review of History, Science and Technology.** The report to the proceedings is included as No. 8 among the "Contributions to the history of astrometry ".
++++++++++++++++++++++++++++++++++++++++++++++++++++++++++++

Further installments in preparation: On the Hipparcos mission studies 1975-79 and on the Hipparcos archives.

*Best regards    Erik*          *http://www.astro.ku.dk/~erik*

# Reports from 2008 and 2011 on History of Astrometry:

Overview, summary and link to individual reports from 2008 and 2011 are placed in an index file which will give direct quick connection to individual reports: www.astro.ku.dk/~erik/1104-index.pdf .

The two collections of reports are placed in two big files at the following links, including overview and summary pages:

The reports from 2008 are placed at arXiv and in a file printing on 8+94 pages: www.astro.ku.dk/~erik/HistoryAll.pdf   and the title is:
"Astrometry and optics during the past 2000 years"

The reports from 2011 are placed at arXiv and in a file printing on 8+46 pages: www.astro.ku.dk/~erik/History2All.pdf   and the title is:
"Astrometry during the past 100 years"